\documentclass[oneside,11pt, showpacs]{revtex4}
\usepackage[T1]{fontenc}
\usepackage[cp1250]{inputenc}
\usepackage{times}
\usepackage[figuresright]{rotating}
\usepackage{psfrag}
\usepackage{amsmath}
\usepackage{latexsym}
\usepackage[hang,small]{caption2}
\usepackage{natbib}

\begin{document}
\title[]{Curie temperature control by band parameters tuning in \\

P\lowercase{b}$_{1-\lowercase{x}-\lowercase{y}-\lowercase{z}}$M\lowercase{n}
$_{\lowercase{x}}$S\lowercase{n}$_{\lowercase{y}}
$E\lowercase{u}$_{\lowercase{z}}$T\lowercase{e}}

\author{I Kuryliszyn-Kudelska$^1$, W Dobrowolski$^1$,
M Arciszewska$^1$,\\ V Domukhovski$^1$, V K Dugaev$^2$$^{,}$$^3$,
V E Slyn'ko$^2$, E I Slyn'ko$^2$ and I M Fita$^1$$^{,}$$^4$}

\address{$^1$ Institute of Physics, Polish Academy of Sciences,\\ Al. Lotnikow 32/46, 02-668 Warsaw, Poland}
\address{$^2$ Institute of Material Science Problems, Ukrainian Academy of Sciences,\\ 5 Wilde Street, 274001, Chernovtsy, Ukraine
}
\address{$^3$ Department of Physics and Center for Physics of Fundamental Interactions,\\ Av. Rovisco Pais, 1049-001 Lisbon, Portugal
}
\address{$^4$ Donetsk Institute for Physics and Technology, National Academy of
Sciences,\\
   R. Luxemburg str. 72, 83114 Donetsk, Ukraine}

\begin{abstract}
We present the study of magnetic and transport properties of
Pb$_{1-x-y-z}$Mn$_{x}$Sn$_{y}$Eu$_{z}$Te. AC magnetic
susceptibility measurements as well as transport characterization
were performed. The obtained results indicate that the presence of
two types of magnetic ions influences magnetic properties of
investigated IV--VI semimagnetic semiconductor. A qualitative
analysis and possible mechanisms of the substantial dependence of
the Curie temperature on the Eu content are presented. The most
likely reason of the observed Curie temperature behaviour is a
strong dependence of the location of a heavy mass $\Sigma$ band
upon the alloy composition. The theoretical calculations in frame
of simple models confirm this point.
\end{abstract}
\pacs{75.50.Pp, 75.30.Kz, 81.40.Rs}  \maketitle

\section{Introduction}

Manipulation of the spin degree of freedom in semiconductors has
become a focus of interest in recent years. In the context of spin
electronics particularly interesting are ferromagnetic and
semimagnetic (diluted magnetic) semiconductors (SMSC's).
Understanding of the carrier mediated ferromagnetism was initiated
by a study of ferromagnetism in IV-VI based SMSC's. In this class
of materials, deviations from stoichiometry result in the carrier
density sufficiently high to produce strong ferromagnetic
interactions between the localized spins. It was shown that in
Pb$_{1-x-y}$Mn$_{x}$Sn$_{y}$Te mixed crystals with high Sn
concentration, $y \geq 0.6$, the free hole concentration can be
varied by means of isothermal annealing in the range between
10$^{20}$ $\div$ 10$^{21}$ cm$^{-3}$ \cite{SGFW86}.
\\
In the present paper, we consider the effect of the presence of
two types of magnetic ions incorporated into semiconductor matrix
on magnetic properties of resultant semimagnetic semiconductor. In
order to simplify the theoretical description of the investigated
magnetic system, two types of magnetic ions were chosen with
spin-only ground state: substitutional Mn$^{2+}$ possesses
$S=5/2$, while Eu$^{2+}$, the second type of magnetic ion, has
$S=7/2$. There are several reasons for which the magnetic
semiconductors based on lead chalcogenides are ideal materials for
such kind of investigations. First, a variety of magnetic
properties has been observed in Mn-based IV-VI SMSC's. Second, the
characteristic feature are semi-metallic electric properties with
the well developed methods of control of carrier concentration.
Pb$_{1-x-y-z}$Mn$_{x}$Sn$_{y}$Eu$_{z}$Te is a unique system in
which the interplay between magnetic and electronic properties can
be observed and studied. In particular, the carrier induced
paramagnet-ferromagnet as well as the ferromagnet-spin glass
transitions have been observed \cite{SGFW86}, \cite{JSSE92}. This
is due to the combination of an RKKY type of interaction between
the magnetic ions with a possibility to manipulate the free
carrier concentration. These two features give IV-VI semimagnetic
materials the distinguished position within the whole family of
semimagnetic semiconductors. The additional advantage is that for
the Pb$_{1-x-y}$Mn$_x$Sn$_y$Te crystals, the parameters of the
energy structure are very well known.

All known Eu-based IV-VI semimagnetic lead chalcogenides with the
carrier density (electrons as well as holes) below 10$^{19}$
cm$^{-3}$ are paramagnetic down to the temperature \emph{T}=1 K
(similarly to Mn-based compounds). A strongly localized character
of 4f orbitals of the rare earth ions results in very weak
exchange interactions both between the magnetic ions and between
ions and the free carriers. As a result, the Sn$_{1-z}$Eu$_{z}$Te
crystals are not ferromagnetic. The reason of a lack of
ferromagnetism in this material is the very small magnitude of
sp--f exchange integral.

The objective of this work was to perform the systematic
measurements of magnetic AC susceptibility as well as transport
characterization of bulk samples of
Pb$_{1-x-y-z}$Mn$_{x}$Sn$_{y}$Eu$_{z}$Te multinary alloy. The
as-grown as well as annealed samples with different concentrations
$x$ of Mn as well as $z$ of Eu were investigated. The
paramagnet-ferromagnet as well as ferromagnet-spin glass
transitions were observed and studied.

\section{Sample preparation and characterization}

The crystals of Pb$_{1-x-y-z}$Mn$_{x}$Sn$_{y}$Eu$_{z}$Te were
grown by the modified Bridgman method. In the present work, the
samples coming from several technological processes were
investigated. The chemical composition of the samples was
determined by X-ray dispersive fluorescence analysis technique
with uncertainty of 10$\%$. Typically, the crystals were cut
crosswise the growth axis to the 1--2 mm thick slices. The
variation of chemical composition along this area is very small
(1--2$\%$). The results of chemical analysis of all investigated
Pb$_{1-x-y-z}$Mn$_{x}$Sn$_{y}$Eu$_{z}$Te samples are gathered in
Table~\ref{table:IV-1}.

The standard powder X-ray measurements revealed that the
investigated samples are single-phase and crystallize in NaCl
structure, similarly as nonmagnetic matrix and
Pb$_{1-x-y}$Mn$_{x}$Sn$_{y}$Te semimagnetic semiconductor. The
measured values of lattice constants for several
Pb$_{1-x-y-z}$Mn$_{x}$Sn$_{y}$Eu$_{z}$Te samples are collected in
Table~\ref{table:a_PSMET}. For comparison, the values of
calculated lattice constant for Pb$_{1-x-y}$Mn$_{x}$Sn$_{y}$Te
crystals with analogous content of Mn and Sn  are also presented
\cite{DMKS94}. The introduction of Mn ions into the nonmagnetic
matrix of Pb$_{1-x}$Sn$_{x}$Te leads to a decrease of the lattice
constant of resultant Pb$_{1-x-y}$Mn$_{x}$Sn$_{y}$Te \cite{S88}.
The careful inspection of Table~\ref{table:a_PSMET} shows that
introduction of Eu ions to Pb$_{1-x-y}$Mn$_{x}$Sn$_{y}$Te lattice
leads to an increase of the lattice constant of resultant
compound.
 All the investigated
Pb$_{1-x-y-z}$Mn$_{x}$Sn$_{y}$Eu$_{z}$Te samples were
characterized by means of low magnetic field transport
measurements. The aim of the transport characterization was to
obtain information about the elementary electric properties of the
investigated samples: the type and density of free carriers and
their mobility. In the case of IV-VI semimagnetic semiconductors,
the carrier concentration is an important parameter since the
change of the concentration influences the magnetic behaviour of
the material.

The Hall bar samples with typical dimensions of 8mm $\times$ 2mm
$\times$ 1mm were used for the transport measurements. The Hall
voltage $V_{\rm H}$ as well as conductivity voltage $V_{\rm
\sigma}$ were measured. The electrical contacts were prepared
always in the same way. First the surface of the specimens was
etched using the solution of Br$_{2}$ and HBr in the proportion
1:20. Next, the gold contacts were deposited by use of gold
chloride water solution on the polished surface of the samples.
Finally, the electrical contacts were made using indium solder and
gold wires. Typical resistance of the samples was equal to about 1
m$\Omega$. This allowed to apply relatively large current (up to
300 mA). The Hall as well as conductivity measurements were
performed at the room and liquid nitrogen temperature. The
standard DC six probe technique at the static magnetic field up to
1T were used. In the present paper, only a nominal Hall
concentration $p$ was determined at the room and liquid nitrogen
temperature. All the investigated samples occurred to be $p$ type
with the high and almost temperature-independent hole
concentration (in the range between 2$\times $10$^{18}$ cm$^{-3}$
and 2$\times $10$^{21}$ cm$^{-3}$). The obtained values of free
hole concentration obtained at room temperature are gathered in
Table~\ref{table:IV-1}. Typical values of mobility were in the
range of a dozen and a few dozen of cm$^2$/Vs. The values of
carrier concentration, conductivity and mobility were determined
with the uncertainty of 15$\%$ at the room temperature and 30$\%$
at the liquid nitrogen temperature.

\section{Magnetic investigations}

In this section, the results of magnetic studies of
Pb$_{1-x-y-z}$Mn$_{x}$Sn$_{y}$Eu$_{z}$Te samples are presented. AC
magnetic susceptibility studies in the temperature range 1.3-150 K
using a mutual inductance method were carried out. The
susceptibility measurements were carried out in AC magnetic field
of frequency range 7-10000 Hz and the amplitude not exceeding 5
Oe.

Generally, in the range of high temperatures all IV-VI
semimagnetic semiconductors are Curie-Weiss paramagnets  with the
temperature dependence of the magnetic susceptibility described by
the Curie-Weiss law:
\begin{equation}\label{C-W}
\chi(T)=C/(T-\Theta),
\end{equation}
where $C=g^{2} \mu _{\rm B} ^{2} S(S+1)N_{\rm M}$ is the Curie
constant
 and $k_{\rm B} \Theta=(1/3)\, S(S+1)x \sum z_{\rm i} I(R_{\rm i})$ is the
paramagnetic Curie temperature (Curie-Weiss temperature). Here,
$N_{\rm M}$ is the concentration of magnetic ions, $z_{\rm i}$ the
number of magnetic neighbors on the $i$th crystallographic shell,
$I(R_{\rm i})$ the exchange integral between the central ion and
its $i$th magnetic neighbors, $S$ the spin of the magnetic ion,
$g$ the spin-splitting $g$-factor, $k_{\rm B}$ the Boltzman
constant, and $\mu_{\rm B}$ the Bohr magneton.

For all the investigated samples the high temperature behaviour of
the inverse low-field susceptibility $\chi ^{-1}$ was nearly
linear and all data fit well to the Curie-Weiss law of the form:
\begin{equation}\label{CW1}
 \chi(T)=C/(T-\Theta)+\chi_{\rm dia}
\end{equation}
where $\chi_{\rm dia}$ is the susceptibility of the host lattice
(all IV-VI semiconductors without magnetic ions are standard
diamagnetic materials with the magnetic susceptibility around
$\chi_{\rm dia}$ $\simeq -3\times $10$^{-7}$ emu/g).

\begin{figure}[h]
\begin{center}
\includegraphics*[angle=0, width=3in]{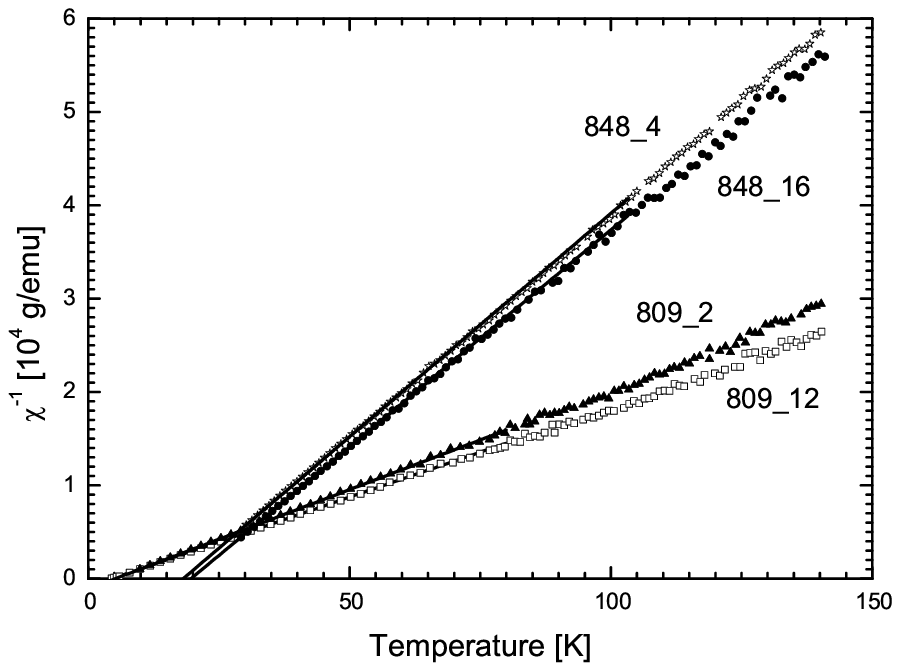}
\caption{The high temperature inverse AC susceptibility measured
for several Pb$_{1-x-y-z}$Mn$_{x}$Sn$_{y}$Eu$_{z}$Te and
Sn$_{1-x-z}$Mn$_{x}$Eu$_{z}$Te samples (809$_{-}$2: $x$=0.031,
$y$=0.850, $z$=0.003; 809$_{-}$12: $x$=0.022, $y$=0.760,
$z$=0.003; 848$_{-}$4: $x$=0.061, $y$=0.927, $z$=0.0115;
848$_{-}$16: $x$=0.050, $y$=0.939, $z$=0.011). The solid lines
correspond to Curie -Weiss law fits. \label{PSMET_1}}
\end{center}
\end{figure}

In the case of PbMnEuTe crystals, the obtained negative and
relatively small values of the paramagnetic Curie-Weiss
temperature $\Theta$ indicate that a weak antiferromagnetic
superexchange interaction is a dominant mechanism of the
interaction. These results correspond to those reported earlier
for PbMnTe. The determined values of paramagnetic Curie
temperature $\Theta$ and Curie constant $C$ are presented in
Table~\ref{table:IV-1}. The inspection of Table~\ref{table:IV-1}
shows that no distinct trends in $\Theta$ and $C$ dependence on Eu
concentration can be observed. However, it should be stressed that
these values are determined with a rather large inaccuracy related
to the uncertainty of the chemical compositions of the samples.
Figure~\ref{PSMET_1} presents the high temperature part of the
inverse AC susceptibility for several PbMnEuSnTe and SnMnEuTe
samples. The fitting procedure (the Curie--Weiss law) revealed the
positive values of the Curie--Weiss temperature in this group of
IV-VI mixed crystals. This indicates the presence of a
ferromagnetic interaction. The obtained values of Curie--Weiss
temperature $\Theta$ and Curie constant $C$ are shown in
Table~\ref{table:IV-1}. The careful inspection of
Table~\ref{table:IV-1} allows to notice significant changes of
Curie-Weiss temperature with the Eu content. The decrease of the
paramagnetic Curie temperature $\Theta$ with the increase of Eu
concentration is clearly visible. The three samples of
Pb$_{1-x-y-z}$Mn$_{x}$Sn$_{y}$Eu$_{z}$Te: 809$_{-}$12,
809$_{-}$30, 809$_{-}$34 are characterized with very similar
values of Mn content and concentration of free holes (see
Table~\ref{table:IV-1}). For the Mn concentration equal to around
$x$ $\simeq$0.02 and free hole concentration $p$=4$\times
$10$^{20}$ cm$^{-3}$ increase of Eu content from $z$=0.003 to
$z$=0.01 leads to the decrease of Curie-Weiss temperature from
4.55 K to 3.02 K and for $z$=0.017 paramagnetic Curie temperature
is equal to 2.63 K. In the case of Sn$_{1-x-z}$Mn$_{x}$Eu$_{z}$Te
crystals such distinct tendency is not observed (see Table
~\ref{table:IV-1}). However, one needs to realize that obtained
values of chemical composition as well as free carrier
concentration are determined with quite large uncertainty.

\begin{figure}[h]
\begin{center}
\includegraphics*[angle=0, width=3.5in]{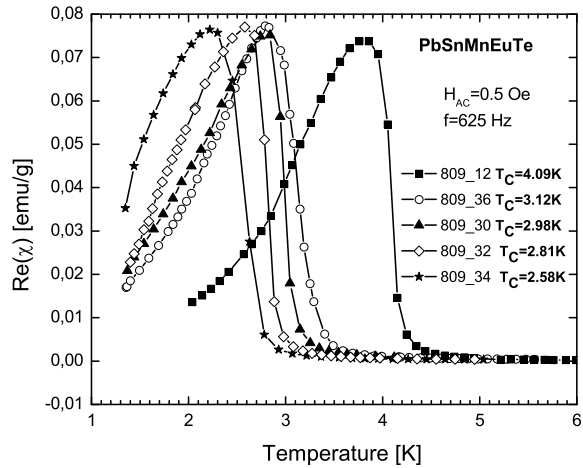}
\caption{The low temperature behaviour of real part of
susceptibility for several
Pb$_{1-x-y-z}$Mn$_{x}$Sn$_{y}$Eu$_{z}$Te samples (809$_{-}$12:
$x$=0.022, $y$=0.760, $z$=0.003; 809$_{-}$36: $x$=0.025, $y$=0.690
$z$=0.013; 809$_{-}$30: $x$=0.024, $y$=0.710, $z$=0.010 ;
809$_{-}$32: $x$=0.026, $y$=0.690, $z$=0.014; 809$_{-}$34:
$x$=0.027, $y$=0.680, $z$=0.017). \label{PSMET_2}}
\end{center}
\end{figure}

\begin{figure}[h]
\begin{center}
\includegraphics*[angle=0, width=3.5in]{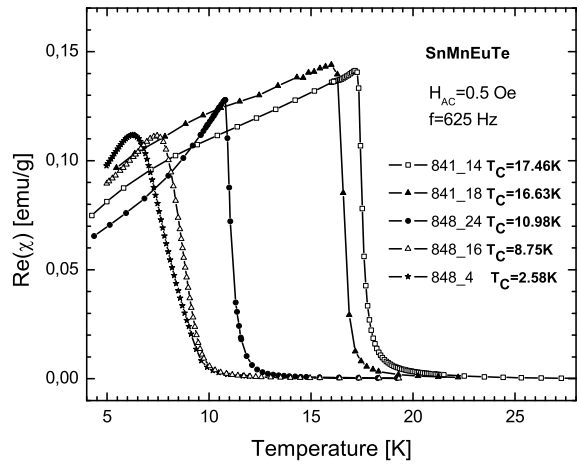}
\caption{The low temperature behaviour of real part of
susceptibility for several Sn$_{1-x-y}$Mn$_{x}$Eu$_{y}$Te samples
(841$_{-}$14: $x$=0.121, $y$=0.873, $z$=0.011; 841$_{-}$18:
$x$=0.128, $y$=0.856, $z$=0.014; 848$_{-}$24: $x$=0.055,
$y$=0.930, $z$=0.0175; 848$_{-}$16: $x$=0.054, $y$=0.939,
$z$=0.011; 848$_{-}$4: $x$=0.058, $y$=0.927, $z$=0.011).
\label{SMET_2}}
\end{center}
\end{figure}

The low temperature studies revealed the presence of
paramagnet--ferromagnet phase transition in the case of SnMnEuTe
as well as PbSnMnEuTe samples. Figure~\ref{PSMET_2} and
Figure~\ref{SMET_2} show the low temperature behaviour of real
component of AC susceptibility Re$(\chi)$ for several samples of
studied Pb$_{1-x-y-z}$Mn$_{x}$Sn$_{y}$Eu$_{z}$Te and
Sn$_{1-x-z}$Mn$_{x}$Eu$_{z}$Te mixed crystals. A typical behaviour
of a ferromagnet is observed. Both real and imaginary components
of the susceptibility dramatically increase at the Curie
temperature $T_{\rm C}$. The Curie temperature was determined by
the maximum slope of $d\,$Re$(\chi)/dT$. The values of $T_{\rm C}$
are approximately equal to the Curie-Weiss temperature $\Theta$
determined from the high temperature susceptibility measurements.
The low temperature measurements confirmed the above-described
tendency for studied Pb$_{1-x-y-z}$Mn$_{x}$Sn$_{y}$Eu$_{z}$Te
crystals, i.e., the decrease of Curie temperature with the Eu
content.

\begin{figure}[h]
\begin{center}
\includegraphics*[angle=0, width=3.5in]{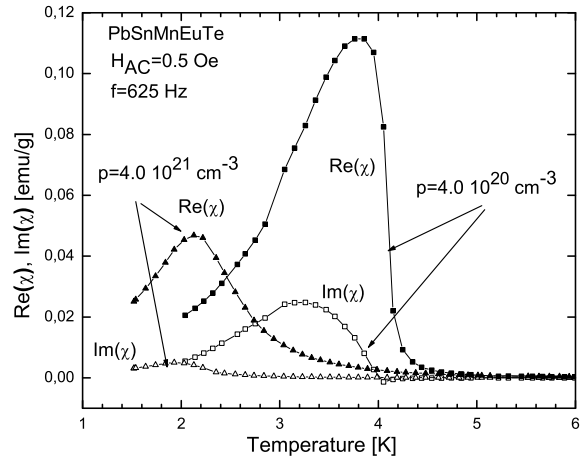}
\caption{The low temperature behaviour of both real $Re(\chi)$ and
imaginary $Im(\chi)$ components of susceptibility for two samples
of Pb$_{1-x-y-z}$Mn$_{x}$Sn$_{y}$Eu$_{z}$Te: 809$_{-}$2
($x$=0.031, $y$=0.850, $z$=0.003, $p=1\times 10^{21}$~cm$^{-3}$)
and 809$_{-}$12 ($x$=0.022, $y$=0.760, $z$=0.003, $p=4\times
10^{20}$~cm$^{-3}$). The typical ferromagnetic characteristics is
observed for 809$_{-}$12 sample and spin glass behaviour for the
sample with higher free hole concentration. \label{PSMET_3}}
\end{center}
\end{figure}

\begin{figure}[h]
\begin{center}
\includegraphics*[angle=0, width=3.5in]{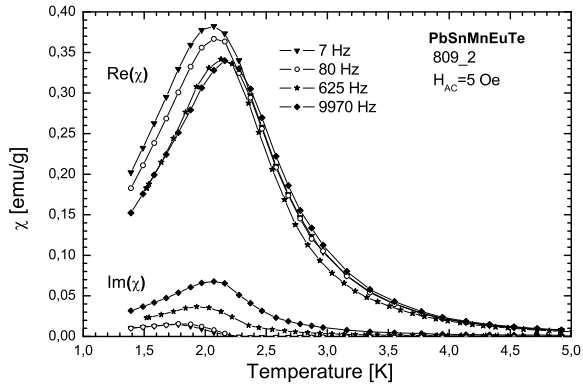}
\caption{The frequency dependence of real Re$(\chi)$ and imaginary
Im$(\chi)$ components of susceptibility for the sample of
Pb$_{1-x-y-z}$Mn$_{x}$Sn$_{y}$Eu$_{z}$Te: 809$_{-}$2: $x$=0.031,
$y$=0.850, $z$=0.003, $p=1\times 10^{21}$~cm$^{-3}$. The shift of
the freezing temperature $T_{f}$ towards higher temperatures with
the frequency increase is clearly visible. \label{PSMET_4}}
\end{center}
\end{figure}

For the sample 809$_{-}$2 of PbSnMnEuTe, the ferromagnet to spin
glass phase transition is observed. Figure \ref{PSMET_3} presents
the characteristic behaviour of low temperature parts of the real
and imaginary components of the susceptibility for the spin glass
(809$_{-}$2) as well as for the ferromagnetic (809$_{-}$12)
samples of Pb$_{1-x-y-z}$Mn$_{x}$Sn$_{y}$Eu$_{z}$Te. In the case
of ferromagnetic sample with the concentration of free holes equal
to $4.0\times 10^{20}$ cm$^{-3}$, the sharp transitions in both
real and imaginary components of susceptibility occur. For the
spin glass sample characterized by higher free hole concentration
$p=1\times 10^{21}$~cm$^{-3}$, a cusp in Re$(\chi)$ is visible at
the freezing temperature $T_{f}$. The magnitude of the
susceptibility at this cusp is much lower than the susceptibility
of the ferromagnetic sample. A corresponding maximum in the out of
phase of susceptibility Im$(\chi)$ is observed at slightly lower
temperature. The 809$_{-}$2 PbMnEuSnTe sample shows an obvious
characteristics of the spin glass--like phase. The cusp observed
in the susceptibility $\chi$ versus temperature $T$ shifts to
higher temperatures when the frequency $f$ of the applied AC field
is increased. This feature -- the increase of the freezing
temperature when the frequency is higher -- was observed in many
well-known canonical spin glass systems \cite{M93}, \cite{T80},
\cite{MDM81}, \cite{HDNM86}. The increase of $T_{\rm f}$ per
decade of frequency is approximately constant, and dependence on
frequency occurs in both real and imaginary parts of AC magnetic
susceptibility. Figure \ref{PSMET_4} presents the frequency
dependence of low temperature parts of real and imaginary
components of susceptibility for 809$_{-}$2 sample of PbMnEuSnTe.
The relative shift of freezing temperature $T_{\rm f}$ per decade
of frequency $R=(\Delta T_{\rm f}/T_{\rm f})\, \Delta \log f$ is
equal to 0.021. The rate of the changes corresponding to maximum
in imaginary part of susceptibility is higher: $R=0.048$.

The values of $R$ reported for known spin glass systems range from
0.005 (Cu) to 0.11 (La$_{1-x}$Gd$_{x}$Al$_{2}$ \cite{DSCN85}) and
the rate of the change in Im$(\chi)$ is the same as the rate of
the change in Re$(\chi)$. The values of $R$ reported for
Sn$_{1-x}$Mn$_{x}$Te are equal: $R$=0.027 ($x$=0.04) \cite{E94},
$R$=0.022 ($x$=0.008) \cite{GTME84}, $R$=0.027 ($x$=0.022)
\cite{GTME84}. It appears that in the case of Sn$_{1-x}$Mn$_{x}$Te
mixed crystals the character of the spin glass phase does not
depend on the manganese concentration. In the case of studied here
Pb$_{1-x-y-z}$Mn$_{x}$Sn$_{y}$Eu$_{z}$Te mixed crystals,
comparable with Mn-based IV-VI magnetic semiconductors, the
significant difference is visible in the inequality of frequency
shift in Re$(\chi)$ and Im$(\chi)$. It has to be noted that
$Im(\chi)$ in conducting media is distorted because of the eddy
currents induced by AC magnetic field. Nevertheless, the obtained
values differ from those obtained for analogous materials
(Sn$_{1-x}$Mn$_{x}$Te), in particular the difference in the rate
of frequency shift of cusp in real and imaginary part of
susceptibility seems to be significant.

\section{Qualitative analysis}

We think that the most likely reason of the strong dependence of
Curie temperature on the Eu content is a variation of the band
structure parameters with the alloy composition. In the analysis
of this problem the following points should be considered.

First, the Eu atom in PbMnSnTe matrix is a magnetic impurity with
spin-only ground state: Eu$^{2+}$ has $S=7/2$. The electrons of
the half-filled f-shell responsible for the moment are very weakly
coupled to the band electrons. The coupling constant $J_{\rm s-f}$
is much smaller than $J_{\rm s-d}$ interaction. Thus, the indirect
RKKY coupling between Eu ions is negligibly small. In addition,
the Eu ion does not feel well any possible magnetization of the
electronic system. As a result, one can not expect a substantial
contribution of Eu magnetic ions to the average magnetization.
Nevertheless, we should not totally exclude a small contribution
from the s-f coupling between Eu atoms and carriers. Obviously, it
would lead to a weak increase of the Curie temperature with the Eu
content. Since the experiment shows the opposite behaviour, we can
assume that the role of Eu as a magnetic impurity is negligibly
small.

On the other hand, EuTe is known to be antiferromagnet. Thus, one
can expect a transition from positive (Curie) to negative (N\'eel)
temperature of magnetic ordering of
Pb$_{1-x-y-z}$Mn$_{x}$Sn$_{y}$Eu$_{z}$Te alloy when changing the
Eu content $z$ from 0 to 1. However, the Eu-Eu interaction is not
mediated by the free carriers, and, therefore, it is short ranged.
For a small Eu content it leads to the antiferromagnetic ordering
for a very small number of Eu-Eu pairs, which do not affect the
interaction mechanism between the Mn ions. If some of the Eu-Eu
neighboring pairs are ordered antiferromagnetically, we can only
expect that they do not contribute to the susceptibility, which is
mostly related to the ferromagnetic Mn-Mn interaction. We conclude
that the suppression of Curie temperature by a possible
antiferromagnetic ordering of Eu-Eu pairs looks very unlikely.

The incorporation of Eu atoms can affect the properties of IV-VI
compounds in a different way, which is not directly related to the
magnetism of Eu ions. Eu is a component of the complex PbMnEuSnTe
alloy, and one should consider a variation of the band parameters
as a function of Eu content. For a qualitative analysis, let us
assume here a simplified two-band model described by some
phenomenological parameters. Note that the real energy spectrum of
IV-VI compounds is rather complicated and one should account for
the nonparabolicity and anisotropy of the energy bands. We use the
following formula describing the indirect RKKY interaction between
Mn ions, when the free carriers from several energy valleys are
taken into account \cite{ESSL95}.
\begin{equation}\label{JRKKY}
J_{\rm RKKY}(R_{\rm ij}) =N\frac{m^{\ast}J_{\rm sd}^{2}a_{\rm
0}^{6}k_{\rm F}^{4}}{32\pi^{3}\hbar^{2}}\; \frac{\sin (2k_{\rm
F}R_{\rm ij})-2k_{\rm F}R_{\rm ij}\cos (2k_{\rm F}R_{\rm
ij})}{(2k_{\rm F}R_{\rm ij})^{4}}
\end{equation}
where $k_{\rm F}$ is the Fermi wave number, $m^{\ast}$ the
effective mass of the carriers, $J_{\rm sd}$ the Mn ion-electron
exchange integral, $a_{\rm 0}$ the lattice constant, \emph{N} the
number of valleys of the valence band, and $R_{\rm ij}$ is the
distance between the magnetic ions. If we take $R_{\rm ij}$ equal
to the mean distance $R$ between the magnetic ions (Mn), this
formula gives the mean interaction energy between magnetic
impurities which is roughly equal to the transition temperature
$T_{\rm C} \approx J_{\rm RKKY}(R)/k_{\rm B}$.

The energy spectrum in the L-points of Brilloin zone can be
described by the Dirac model
\begin{equation}\label{E_L}
  E_{\rm L}(k)=(\Delta^{2}+v^{2}k^{2})^{1/2},
\end{equation}
where $v$ is the band-coupling constant, for which we assume
$v=5\times10^{-8}$~eV \cite{LB82}, $\Delta = E_{\rm g}/2$, and
$E_{\rm g}$ is the energy gap depending on the alloy composition.

There is another "heavy-hole" band with 12 minima in $\Sigma $
points of the Brilloin zone. The energy spectrum of this band can
be approximated as parabolic, and situated at the energy distance
$\epsilon _0$ from the bands described by equation (\ref{E_L})
\begin{equation}\label{E_S}
  E_{\Sigma}(k)=\epsilon_{0}+\frac{\hbar^{2}k^{2}}{2m^{\ast}_{\Sigma}},
\end{equation}
where $m^{\ast}_{\Sigma}\simeq 3m_{0}$, and $m_{0}$ is the free
electron mass. It is commonly known that the energy spectrum of
Pb$_{1-z}$Eu$_{z}$Te alloy is very sensitive to the concentration
of Eu. The energy gap $E_{\rm g}$ depends strongly on Eu content
-- ranging from 189.7~meV for $z=0$ (PbTe) to 248~meV for
$z=0.013$ at the temperature $T=10$ K \cite{LB82}, $d E_{\rm
g}/dz=5.788$~eV at $T=10$ K for $z<0.05$ \cite{S??}. Considering
the energy gap $E_{\rm g}$ dependence on Sn content $z$ of
Pb$_{1-x-y}$Mn$_{x}$Sn$_{y}$Te alloy \cite{LB82,S??}, the
following formula describing the energy gap dependence on the
alloy composition in Pb$_{1-x-y-z}$Mn$_{x}$Sn$_{y}$Eu$_{z}$Te
crystals can be assumed:
\begin{equation}\label{Eg}
 E_{\rm g}=0.19\, (1-y)-0.3\, y+5.788\, z \hspace{0.25cm} [\rm eV]
\end{equation}
In the present calculations the Mn concentration $x$ was accepted
as equal to 0.02.

We also assume \cite{LB82}, \cite{NS98} that the composition
dependence of $\epsilon_{0}$ parameter has the following form:
\begin{equation}\label{eps}
  \epsilon_{0}=0.8\, (1-y)+0.3\, y+8\, z \hspace{0.25cm} [\rm eV]
\end{equation}
Using Eqs.~\ref{E_L} and ~\ref{E_S} the hole concentration in L as
well as in $\Sigma$ valley can be found:
\begin{equation}\label{p_L}
  p_{\rm L}=\frac{(E_{\rm F}^{2}-\Delta^{2})^{3/2}}{3\pi v^{3}},
\end{equation}

\begin{equation}\label{p_S}
  p_{\Sigma}=\frac{1}{3 \pi ^{2}}
  \left[ \frac{2m^{\ast} _{\Sigma}}{\hbar^{2}}(E_{\rm F}-\epsilon_{0})\right] ^{3/2}.
\end{equation}
The Fermi energy $E_{\rm F}$ is determined by the equation:
\begin{equation}\label{pp}
  4p_{\rm L}+12p_{\Sigma}=p_{0},
\end{equation}
which gives the total hole concentration $p_{0}$ and takes into
account the degeneracy of each valley. It is accepted here that
$p_{0}$ is equal to $4\times10^{20}$~cm$^{-3}$.

 The Fermi momentum in
the $\Sigma$ band $k_{\rm F}^{\Sigma}=[2m^{\ast}_{\Sigma}(E_{\rm
F}-\epsilon_{0})]^{3/2}$ can be found as a solution of
Eqs.~(8)-(10). Next, Eq.~\ref{JRKKY} with $k_{\rm F}=k_{\rm
F}^{\Sigma}$, $R=(3a_{0}^{3}/4\pi x)^{1/3}$, $a_{0}=6.5\times
10^{-8}$ cm, $N=12$, $J_{\rm sd}=1$~eV is used for $T_{\rm C}$
calculations. It implies that the main contribution to the RKKY
interaction between Mn ions is related to the heavy holes in
$\Sigma $ bands.

\begin{figure}[h]
\begin{center}
\includegraphics*[width=4in, angle=0]{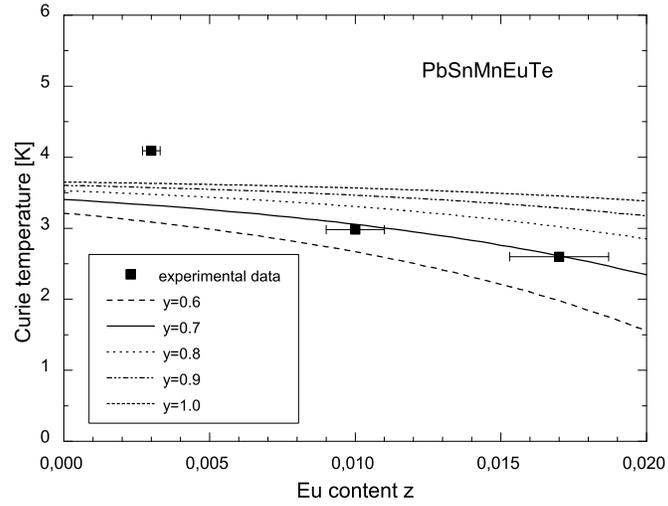}
\caption{Comparing of experimentally determined dependence of Curie
temperature on Eu concentration z (for
Pb$_{1-x-y-z}$Mn$_{x}$Sn$_{y}$Eu$_{z}$Te samples with similar Mn
content $x$ $\simeq$0.02 and free hole concentration
$p$$\simeq$4$\times $10$^{20}$ cm$^{-3}$) with the results of simple
two--band model calculations. The calculated Curie temperature as a
function of Eu content $z$ for various values of Sn concentration
0.6 $\leq y \leq$ 1 and Mn concentration $x$=0.02 was divided by
factor 1.95. \label{T_C}}
\end{center}
\end{figure}

The obtained $T_{\rm C}$ dependence on Eu concentration $z$ for
various values of Sn content ($0.6 \leq z \leq 1$) is shown in
Fig.~\ref{T_C}. The experimental data (Curie temperature
determined for samples with similar Mn content $x$ $\simeq$0.02
and free hole concentration $p$$\simeq$4$\times $10$^{20}$
cm$^{-3}$) as well as calculated Curie temperature values divided
by factor 1.95 are presented for comparison. As we see, the
results of calculations within a simple two-band model can explain
the observed experimentally tendency of Curie temperature decrease
with Eu content. The obtained values of Curie temperature are
higher than determined experimentally. However, it should be
stressed that not all phenomenological parameters of the model are
known precisely, and some simplifying assumptions are used.
Nevertheless, the calculated dependence of Curie temperature on Eu
content reflects very well the experimentally observed effect of
the $T_{\rm C}$ decrease with Eu concentration $z$.

\section{Summary}

In this paper the results of magnetic and transport studies of
Pb$_{1-x-y-z}$Mn$_{x}$Sn$_{y}$Eu$_{z}$Te multinary alloys are
reported. The following results were obtained.

The presence of two types of magnetic ions (Mn and Eu) in IV-VI
semiconductor matrix influences the magnetic properties of
resultant magnetic semiconductor. The results of magnetic
measurements show that the Curie temperature $T_{\rm C}$ as well
as the Curie--Weiss temperature $\Theta$ decrease with the
increase of Eu content in Pb$_{1-x-y-z}$Mn$_{x}$Sn$_{y}$Eu$_{z}$Te
samples. The magnetic susceptibility measurements revealed also
that Eu changes the spin glass dynamics in this material. The
difference in the rate of frequency shift of cusp in the real and
imaginary parts of susceptibility is visible. Such behaviour was
not observed for Mn-based IV-VI magnetic semiconductors.

Our qualitative analysis shows that a variation of the band
parameters with the alloy composition can be responsible for the
observed strong dependence of the Curie temperature on Eu content.
A simple two band model explains both the order of the transition
temperature values and $T_{\rm C}$ dependence on Eu concentration.
The calculated dependence of Curie temperature on Eu content
reflects very well the experimentally confirmed effect of $T_{\rm
C}$ decrease with Eu concentration $z$.

The observed dependence of the Curie temperature on Eu content
clearly demonstrates that the magnetism of semiconductors can be
effectively controlled by using an energy band structure
dependence on alloy composition. The situation with several
non-equivalent energy minima is not unique for IV-VI
semiconductors. The similar effect exists also in Ge-Si alloys and
in other semiconductors (like high-energy minima in GaAs). We
believe that it can be used to increase the critical temperature
of ferromagnetism in magnetically-doped semiconductors.

\section{Acknowledgments}
This work is partly supported by FCT Grant No.~POCI/FIS/58746/2004
in Portugal and by the Polish State Committee for Scientific
Research under Grant No.~2~P03B~053~25.

\newpage
\addcontentsline{toc}{chapter}{Bibliography}
\bibliographystyle{unsrt}
\bibliography{Kuryliszyn_Kudelska}

\newpage

\newpage

\begin{table}
\caption{The results of magnetic and tranportmeasurements for
IV-VI mixed crystals. $C$ is the Curie constant, $\Theta$ is the
paramagnetic Curie temperature (Curie-Weiss temperature), $T_{C}$
is the Curie temperature, $T_{f}$ is the freezing temperature, $p$
is the free hole concentration obtained at room temperature.}
\label{table:IV-1}

\begin{tabular}{|c|c|c|c|c|c|c|c|c|c|c|c|}
\hline
sample&$x$&$z$&$y$&$C$&$\Theta$&$T_{C}$&$T_{f}$&$p$\\

number  & & & &[emu/g]&[K]&[K]&[K]&[10$^{21}$ cm$^{-3}]$\\

\multicolumn{9}{c}{Sn$_{1-x-z}$Mn$_{x}$Eu$_{z}$Te}\\

841\_14&0.116&0.011&0.873&0.00215&19.65&17.46&-&1.40\\

841\_18&0.131&0.013&0.856&0.00209&18.18&16.63&-&1.30\\

842\_4&0.063&0.0045&0.932&-&-&10.90&-&1.40\\

842\_8&0.068&0.003&0.929&-&-&13.07&-&1.56\\

842\_14&0.070&0.007&0.923&-&-&15.23&-&1.56\\

842\_20&0.091&0.009&0.900&0.00230&17.98&17.10&-&1.21\\

848\_4&0.061&0.0115&0.927&0.00141&11.57&11.57&-&1.77\\

848\_10&0.064&0.012&0.924&-&-&8.35&-&1.93\\

848\_16&0.050&0.011&0.939&0.00149&11.02&8.75&-&1.24\\

848\_22&0.065&0.018&0.917&-&-&10.75&-&1.37\\

848\_24&0.051&0.019&0.930&0.00222&12.08&10.98&-&1.57\\

848\_26&0.074&0.023&0.903&-&-&11.31&-&1.66\\
\multicolumn{9}{c}{Pb$_{1-x-y-z}$Mn$_{x}$Sn$_{y}$Eu$_{z}$Te}\\

809\_2&0.031&0.003&0.850&0.00047&5.16&-&2.0&1.01\\

809\_4&0.030&0.002&0.850&0.00050&4.88&-&-&0.501\\

809\_10&0.030&0.003&0.780&0.00050&4.74&4.13&-&0.601\\

809\_12&0.022&0.003&0.760&0.00052&4.55&4.09&-&0.401\\

809\_30&0.024&0.010&0.710&0.00080&3.02&2.98&-&0.425\\

809\_30 ann&0.024&0.010&0.710&0.00080&4.16&3.55&-&0.822\\

809\_32&0.026&0.014&0.690&0.00082&2.89&2.81&-&0.325\\

809\_34&0.027&0.017&0.680&0.00090&2.60&2.60&-&0.449\\

809\_36&0.025&0.013&0.690&0.00077&3.13&3.12&-&0.318\\
\multicolumn{9}{c}{Pb$_{1-x-z}$Mn$_{x}$Eu$_{z}$Te}\\

793\_2&0.010&0.000&-&0.00039&-0.93&-&-&-\\

\pagebreak

793\_4&0.010&0.001&-&0.00038&-0.69&-&-&-\\

793\_4&0.010&0.001&-&0.00038&-0.69&-&-&-\\

793\_6&0.009&0.009&-&0.00051&-1.12&-&-&-\\

793\_10&0.005&0.004&-&0.00053&-0.42&-&-&-\\

793\_12&0.007&0.003&-&0.00059&-0.39&-&-&-\\

793\_14&0.005&0.005&-&0.00086&-0.41&-&-&-\\
\hline
\end{tabular}
\end{table}

\begin{table}[tbp]

\caption{The lattice constant \emph{a$_{0}$} of
Pb$_{1-x-y-z}$Mn$_{x}$Sn$_{y}$Eu$_{z}$Te samples determined by the
standard powder X-ray measurements and the values of the
calculated lattice constant of  Pb$ _{1-x-y}$Mn$_{x}$Sn$_{y}$Te
\emph{a} \cite{DMKS94} with similar content of Mn and Sn.
\label{table:a_PSMET}}
\begin{center}
\begin{tabular}{|c|c|c|c|c|c|c|}
\hline \# of the sample& \emph{x}& \emph{z}&
\emph{y}& \emph{a$_0$} & \emph{a} \\
&&&&[\AA]&[\AA]\\
 \hline
809\_2 & 0.031& 0.0027& 0.85& 6.3130& 6.2866\\
809\_4 & 0.030& 0.0016& 0.85& 6.3237& 6.2876\\
809\_12& 0.022& 0.003& 0.76& 6.3375& 6.3113\\
809\_28& 0.020& 0.007& 0.73& 6.3563& 6.3309\\
809\_36& 0.025& 0.013& 0.69& 6.3427& 6.3311\\
\hline
\end{tabular}
\end{center}
\end{table}

\end{document}